# High Temperature Superconductivity – Magnetic Mechanisms

M. R. Norman

Materials Science Division, Argonne National Laboratory, Argonne, IL  60439

**Abstract:**  A brief history is offered concerning the relation of magnetism to superconductivity, and the possibility that magnetic correlations are responsible for certain types of superconductors.  A central focus is on high temperature cuprate superconductivity and the important question of whether its d-wave pairing is caused by antiferromagnetic or singlet correlations.  Connected with this question is the much debated relation of the pseudogap phase to the superconducting phase, and whether lattice degrees of freedom are relevant or not.

# A Brief History – Part 1

The histories of superconductivity and magnetism have been much intertwined. The discovery of the Meissner effect in 1933, where magnetic flux is expelled from a superconductor as it is cooled below its transition temperature $T_c$, demonstrated that superconductors were more than just perfect conductors, leading to the famous proposal by London of the existence of a macroscopic condensate that accounts for the supercurrent. This was quite startling given the fermionic nature of the charge carriers in a metal, whose statistics do not prefer condensate formation as in the case of bosons. This idea was codified by Ginzburg and Landau in 1950, who introduced an order parameter field describing the condensate of electrons. The resulting Ginzburg-Landau Hamiltonian has been exploited in many areas of physics, including those dealing with the origin of the universe. One of the early successes of Ginzburg-Landau theory was the prediction of type II superconductivity by Abrikosov in 1957, where above a lower critical field $H_{c1}$, magnetic flux penetrates in the form of quantized vortices. Recently, such vortices have been observed in cold atom condensates as well.

It was the development of a microscopic theory by Bardeen, Cooper, and Schrieffer in 1957, and the subsequent proof by Gor'kov in 1959 of the equivalence of this theory to the Ginzburg-Landau formalism, which began the modern era of superconductivity (Schrieffer, 1964). The crucial finding of Cooper in 1956 was that in the presence of a Fermi sea, arbitrarily weak attraction could lead to pair formation. Such pairs behave as bosons, thus explaining how a condensate could exist. The attraction that forms the pairs is a consequence of the positive ions and the fact that the ions and electrons have different time scales. Figure 1 illustrates how this works. Positive ions are attracted to negative electrons. This polarizes the ions towards the electron. When the electron leaves, a second electron sees this positive cloud and is attracted to this location, leading to pair formation. As the interaction is local in space, one forms s-wave pairs that are spin singlets by fermion antisymmetry. The pairs form despite the presence of the Coulomb repulsion of the individual electrons as they are at the same place, but at different times. Once the electrons become energetic enough relative to the ion vibrations (phonons), the attraction goes away, and the Coulomb repulsion wins out. This 'retardation' effect is what is responsible for limiting the superconducting transition temperature of electron-phonon systems to values significantly smaller than room temperature.

The development of the BCS theory also led to the understanding of why proximity to magnetism is usually detrimental to superconductivity. In 1960, Abrikosov and Gor'kov showed, using the powerful Matsubara technique for finite temperature quantum field theory, that magnetic impurities are pair breaking (Abrikosov and Gor'kov, 1961). This result is easy to appreciate, since spin flip scattering will destroy singlet pairs. This argument was generalized to the dynamic case by Berk and Schrieffer, who showed by summing a ladder series representing repeated scattering between the two electrons of the pair, that ferromagnetic spin fluctuations were detrimental to spin singlet superconductivity (Berk and Schrieffer, 1966) (Figure 2).

In 1968, though, Fay and Layzer turned this argument around (Fay and Layzer, 1968). The same formalism can be used to show that ferromagnetic spin fluctuations will promote spin triplet pairing, which is p-wave due to fermion anisymmetry. This relies on the fact that the bare contact interaction is zero in the triplet channel (due to the Pauli exclusion principle), but the induced interaction (again, from the ladder sum of repeated scattering of two electrons via spin fluctuations shown in Figure 2) is attractive. Physically, this 'attraction' is due to the fact that an up spin electron prefers to have other up spin electrons nearby. The node in the pair wavefunction for the p-wave case acts to prevent the two electrons from coming too close together, thus minimizing the detrimental effects due to the direct Coulomb repulsion. They suggested that this mechanism could apply to nearly ferromagnetic metals such as Pd and also to the charge neutral case of $^3$He.

In 1972, Osheroff, Richardson, and Lee indeed discovered p-wave superfluidity in $^3$He. It was soon realized that there were two superfluid phases, an anisotropic A phase and an isotropic B phase. This was difficult to understand, since the free energy of the isotropic phase should be the lowest according to Ginzburg-Landau theory. But in the following year, Anderson and Brinkman showed how the anisotropic A phase could be stabilized (Anderson and Brinkman, 1973). The development of an energy gap removes some of the very spin fluctuations that lead to pairing in a spin fluctuation model, thus leading to a decrease in the pairing kernel. This gapping effect is less pronounced in the anisotropic A phase than in the isotropic B phase, explaining how the A phase can be stabilized. This would seem to have been the 'smoking gun' for spin fluctuations, but in subsequent years, it was realized that there are many contributions to the pairing kernel besides spin fluctuations, such as density fluctuations, transverse current fluctuations, etc. In fact, alternates to spin fluctuation theory has been proposed in the case of $^3$He, including the 'nearly localized' approach of Vollhardt and the polarization potential model of Bedell and Pines (Vollhardt and Wolfle, 1990).

At this point, an important issue should be realized. Unlike the electron-phonon case where electrons and ions can be approximately treated as separate systems, the spin fluctuations themselves are composed of electrons. This makes the whole notion of 'pairing glue' suspect in this case. One consequence of this is that spin fluctuations do not appear to obey Migdal's theorem. How this theorem works for the classic phonon case is as follows. The ratio of the electron mass to the ion mass is very small, thus leading to a controlled perturbation expansion. For most cases, it is sufficient to stop at lowest order when evaluating the electron and phonon self-energies. The exception is the pairing instability, which requires summing a ladder series of repeated scattering. This neglect of vertex corrections, though, is not generally valid in the case of spin fluctuations, as shown by Hertz, Levin, and Beal-Monod (Hertz *et al.,* 1976).

In 1979, Frank Steglich's group discovered superconductivity in the heavy electron alloy CeCu$_2$Si$_2$. This flew in the face of existing wisdom that proximity to magnetism was deadly for superconductivity in metals ($^3$He obviously differing in that its pairs are composed of charge neutral atoms). Soon, several of these materials were discovered, and one of them, UPt$_3$, was known to exhibit strong spin fluctuation behavior. As expected, various spin fluctuation theories based on $^3$He were proposed to explain the superconductivity seen in these materials. But subsequent neutron scattering revealed that these metals were nearly antiferromagnetic, rather than nearly ferromagnetic. In 1986, several theories were formulated for heavy fermions based on the neutron scattering data (Miyake *et al.,* 1986; Scalapino *et al.,* 1986). The prediction was singlet pair formation, this time due to the fact that in the presence of antiferromagnetic correlations, an up spin electron prefers to be surrounded by down spin electrons. To avoid the strong on-site Coulomb repulsion, the pair state has d-wave symmetry. For a simple cubic lattice, the pairs take the form $(x^2-y^2) \pm i(3z^2-r^2)$. In real space, this pair wavefunction corresponds to six lobes that point from a given atomic site to its near neighbors.

In hexagonal symmetry, this two dimensional group representation instead becomes isomorphic to L=2, M=±1 spherical harmonic. This $E_{1g}$ model is a leading candidate to describe various experimental data in $UPt_3$, including the observation of the three different superconducting phases seen in the H,T phase diagram (Joynt and Taillefer, 2002). But problems with this model have led to a variety of other proposals, including the A, B model of Garg (two nearly degenerate single dimensional group representations), the p-wave model of Machida, and the f-wave model of Norman and Sauls. This last model ($E_{2u}$) provides a particularly good description of the H,T phase diagram, thermal conductivity, and transverse ultrasound data (Sauls, 1994). Triplet models (p,f) for $UPt_3$ are particularly attractive, since there has been no observed change in the Knight shift (i.e., the static spin susceptibility) when going below $T_c$, as opposed to other heavy fermion superconductors such as $UPd_2Al_3$ and $CeCu_2Si_2$ (Tou *et al.*, 2005). The point to be made, though, is that despite many claims in the literature, the actual pairing symmetry of any heavy electron superconductor is unknown at present.

# A Brief History – Part 2

In early 1986, high temperature superconductivity was discovered by Bednorz and Mueller in the doped perovskite $La_{2-x}Ba_xCuO_4$ (Bednorz and Mueller, 1986). It was not until November of that year, though, before the results was verified and thus led to wide scale recognition. By January of 1987, superconductivity above the temperature at which air liquefies was found in the related compound $YBa_2Cu_3O_7$ by Chu and collaborators. The same month, a theory for these materials was proposed by Anderson (Anderson, 1987). He recognized that the undoped compound $La_2CuO_4$ would likely be a Mott insulator. He speculated that the Néel (antiferromagnetic) order of the insulator would be melted by quantum fluctuations (due to the low spin S=1/2 of the $d^9$ Cu ion and the two dimensional nature of the $CuO_2$ planes). Although subsequently Néel order was discovered, it indeed disappears when only a few percent of holes are doped into the material. Anderson denoted this melted Néel state as a resonating valence bond (RVB) state, which represents a liquid of spin singlet pairs. When the system is doped, the presence of charge carriers causes this spin pairing state to condense into a superconducting state. Originally, it was thought that the resulting pair symmetry would be s-wave like, but subsequent work in 1988 predicted d-wave symmetry instead (Kotliar and Liu, 1988; Zhang *et al.*, 1988).

It was not long after Anderson's theory was announced in 1987 that more traditional spin fluctuation based approaches were brought to bear on this matter. Bickers, Scalapino, and Scalettar observed that the 3D pairing state discussed earlier in the context of heavy fermion materials would reduce to $dx^2-y^2$ in two dimensions (Bickers *et al.*, 1987). Scalapino subsequently gave an intuitive picture of how such a pair state arises (Scalapino, 1995). In momentum space, the zero frequency limit of the real part of the effective potential coming from the Coulomb interaction is repulsive for all wavevectors, with a maximum at a wavevector $(\pi,\pi)$ which would be the ordering wavevector for the undoped antiferromagnet (Figure 3). But Fourier transformed into real space, the effective potential has Friedel oscillations. Although obviously repulsive at short distances, the potential is attractive for near neighbor separations (Figure 3). In momentum space, this is reflected in the gap equation, where the sign change of the d-wave tight binding gap function $\Delta(\mathbf{k}) = \cos(k_x a)-\cos(k_y a)$ upon translation by $(\pi,\pi)$ compensates for the repulsive sign of the potential V: $\Delta(\mathbf{k}) = \Sigma_{\mathbf{k}'} V_{\mathbf{kk}'} \Delta(\mathbf{k}')$ where the pairing kernel $V_{\mathbf{kk}'} = U + J \ (\cos(q_x a)+\cos(q_y a))$ with $\mathbf{q=k-k'}$, U positive and J negative.

Despite this initial success, there were no indications from experiment at that time supporting the existence of a d-wave pairing state. In fact, because the cuprates are somewhat dirty systems replete with impurities, the feeling was that the order parameter would have to be s-wave like to avoid pair breaking. To understand this, we note that the d-wave order parameter changes sign under a reflection operation xy→yx. Any impurity scattering that mixes these two states destroys the d-wave phasing relation.

In the early 1990s, though, experimental evidence began to emerge supporting a d-wave picture. The temperature dependence of the NMR spin relaxation rate, the Knight shift, and the in-plane penetration depth, did not follow the exponential behavior predicted for s-wave pairing, but rather the power law behavior predicted for a d-wave state due to the presence of a node (zero) in the order parameter. This node was subsequently imaged directly by angle resolved photoemission. Then phase sensitive Josephson tunneling saw the predicted sign change of the d-wave order parameter upon 90 degree rotation. Since then, a large body of experimental evidence has accumulated, including the dependence of $T_c$ on impurities, that overwhelming confirms the d-wave nature of the pairs. At the same time, there has been great progress in the theory of spin fluctuations as applied to cuprates, as well as in solving the underlying microscopic theories based on the Hubbard model. In addition, there have been recent advances made in the RVB theories as well.

## Electronic Structure of the Cuprates

In the undoped cuprates, the copper ions are in a $d^9$ configuration. This corresponds to a single hole in the $x^2-y^2$ orbital (not to be confused with the d-wave pair state discussed earlier). Of all transition metal oxides, the cuprates are unusual in that the copper d orbital and the oxygen p orbital have energies that are nearly degenerate (Pickett, 1989). As a consequence, the dominant energy scale in the problem is the large ($\sim 6$ eV) bonding-antibonding splitting between the copper $dx^2-y^2$ orbital and the oxygen $p_x$ and $p_y$ orbitals (Figure 4). This leaves the highest energy band (the antibonding one) as half filled. Adding Coulomb repulsion, this band splits into two, a lower Hubbard band and an upper Hubbard band. The resulting Mott insulating gap is of order 2 eV. Keeping all three bands, this is known as the three band Hubbard model, but keeping just the antibonding band is known as the single band Hubbard model. Almost all treatments assume the latter, though Varma has argued that important physics is thrown out upon such a reduction (as will be discussed later). In the limit of large Coulomb repulsion, U, one can then project onto the subspace which does not allow double occupation of the Cu site, leading to the t-J model, where J, the superexchange interaction, is proportional to $t^2/U$ and t is the effective hopping integral between Cu sites. J prefers antiferromagnetic orientation of the copper spins (one spin per site in the undoped case). This can be seen by the fact that the Pauli exclusion principle does not allow virtual double occupation unless the two spins are anti-aligned.

There are no exact solutions of either the single band Hubbard model or the simpler t-J model in two dimensions. Approximate treatments have been done using quantum Monte Carlo, density matrix renormalization group, dynamical mean field theory (and its various cluster extensions), and exact diagonalization of small clusters. Although there is no solid proof at this time, results are encouraging enough that there is a strong probability that a true d-wave pairing instability exists in these models. It is beyond the scope of this article to review these techniques. Instead, we give an overview of the spin fluctuation approach and its RVB counterpart, discuss these theories in relation to experimental data, and then end with a discussion of alternate mechanisms for cuprate superconductivity.

## Spin Fluctuation Theories

The literature on this subject is vast, so this brief review can only give the highlights. The basic idea is that in spin fluctuation theories, the pair potential V is found to be proportional to $I^2$ Im $\chi(\mathbf{q},\omega)$ where I is the effective spin interaction between electrons and Im $\chi(\mathbf{q},\omega)$ is the imaginary part of the dynamic spin susceptibility, with the proportionality prefactor of order unity (-3/2 for spin singlet pairs and +1/2 for spin triplet pairs). "I" itself is dependent on the underlying theory. For instance, in the Hubbard model, this would be the Hubbard interaction U, but if one used an effective low-energy theory, then it would be J, the superexchange interaction. At this level, the theory is equivalent to RPA (random phase approximation) with $\chi(\mathbf{q},\omega)=\chi_0(\mathbf{q},\omega)/(1-I\chi_0(\mathbf{q},\omega))$ with $\chi_0(\mathbf{q},\omega)$ the polarization bubble calculated using bare Greens functions.

There are several ways to consider going beyond RPA. One is simply to add fluctuation corrections to $\chi(\mathbf{q},\omega)$. To understand this approach (Lonzarich and Taillefer, 1985), we note that in a Ginzburg-Landau expansion for magnetism, the free energy would be of the form $aM^2 + bM^4$ where M is the magnetization (staggered magnetization in the antiferromagnetic case). With fluctuations included, we note that upon factorization of $M^4$ one obtains a term of the form $6<M^2>M^2$ where $<M^2>$ is the expectation value of $M^2$ averaged over all statistical ensembles. This $6bM^2<M^2>$ term then renormalizes the $aM^2$ term, leading to an $a_{eff}=a+6b<M^2>$, noting that "a" is simply the inverse (RPA) susceptibility (in statistical field theory, this is often denoted as the Hartree approximation). This approach has been enormously successful in describing transition metal magnets, for instance, predicting the lack of magnetic long range order in lower dimensions, and understanding why most magnets have transition temperatures strongly suppressed relative to mean field (Stoner) theory. In turn, these fluctuations enter into the pair kernel, and this approach has been extensively studied by Moriya and co-workers in the context of a spin fluctuation mediated picture for cuprate superconductivity (Moriya and Ueda, 2000 and 2003).

Another way to go beyond RPA is to use dressed Greens functions rather than bare ones when constructing the polarization bubble. This is the basis behind the fluctuation exchange (FLEX) approximation (Bickers *et al.*, 1989), where the self-energy used to dress the single particle Greens functions is chosen to satisfy a certain self-consistency relation involving the free-energy, the self-energy, and the Greens function (the conserving approximation of Gordon Baym). Subsequently, this method has been applied by many authors, not only to address the single-particle spectral function and the dynamic spin susceptibility, but also the pairing interaction.

One issue with such approximations is that it is usually dangerous to dress the Greens functions without including vertex corrections in the spin susceptibility. In the "two-particle self-consistent" approach of Vilk and Tremblay (Vilk and Tremblay, 1997), a similar procedure to FLEX is done, but now the interaction U (they assumed a Hubbard model) is replaced by $U_{sp}$, with $U_{sp}$ a screened interaction (again chosen to satisfy certain self-consistent relations), which enters the susceptibility, $\chi(\mathbf{q},\omega)=\chi_0(\mathbf{q},\omega)/(1-U_{sp}\chi_0(\mathbf{q},\omega))$, where as in FLEX $\chi_0$ is calculated using dressed Greens functions. In essence, $U_{sp}$ represents a constant vertex correction. More sophisticated approximations would allow $U_{sp}$ to depend on momentum and frequency. Note that in such approaches, the $U^2$ prefactor of the pairing kernel is now replaced by $UU_{sp}$. A recent review of this and related approaches based on dynamical mean field theory (and its cluster extensions) has been offered by Tremblay, Kyung and Senechal (Tremblay *et al.,* 2006).

One can, of course, attempt to go beyond these approximations by performing a systematic diagrammatic expansion including vertex corrections, for instance the work of Doug Scalapino and co-workers (Bulut, 2002). These authors have used quantum Monte Carlo simulations as a fundamental check of their work (as has the Tremblay group). The limitation is that such quantum Monte Carlo simulations cannot be carried out at low temperatures because of the so-called fermion sign problem (the many-body wavefunction having both positive and negative regions in the fermionic case leads to the problem of negative probabilities in the context of the simulations).

And, there have been some attempts to combine all of these ideas into a single approach. A good example is the extensive work of Chubukov and collaborators on a quantum field theoretical approach to the spin fluctuation problem, dealing with matters concerning the strong influence of quantum and thermal fluctuations, with the resulting non Fermi liquid behavior. Space prohibits an adequate summary of this work, and the reader is referred to a review article this group has done (Abanov *et al.,* 2003).

Finally, one can go the phenomenological route and replace $\chi(\mathbf{q},\omega)$ in the pairing kernel by the experimental dynamic spin susceptibility. This approach has been exploited by a number of authors, in particular David Pines and co-workers (Monthoux and Pines, 1994), where they modeled $\chi(\mathbf{q},\omega)$ based on the NMR data of Slichter's group. Subsequent work has exploited the growing amount of data for $\chi(\mathbf{q},\omega)$ obtained by inelastic neutron scattering.

So, what is the upshot of these approaches. For doping ranges relevant for experiment, they predict $dx^2-y^2$ pairing. The physics is essentially equivalent to that discussed earlier in this article, and in the context of the cuprates, this was first discussed by Bickers *et al* (Bickers *et al.,* 1987) as mentioned before. But what does this all mean when addressing experimental data?

## Spin Fluctuation Theories – Confronting Experiment

First, we deal with the "if" story. That is, does d-wave pairing really emerge from the underlying Hamiltonians (single band Hubbard, t-J) that underlie these spin fluctuation approaches? Interestingly, the jury is still out on this question. The most detailed diagrammatic studies of Scalapino and co-workers (Bulut, 2002) have not definitely answered this question (the issue being whether vertex corrections do or do not suppress the pairing instability). Quantum Monte Carlo simulations have yielded conflicting results, some indicating an enhancement of pairing, others not, though the most recent studies indicate an enhancement (Sorella *et al.*, 2002). The issue, of course, is the inability to access very low temperatures because of the fermion sign problem.

Of course, virtually all such approaches do yield d-wave pairing (within a given approximation), but the predicted values of $T_c$ vary quite a bit. This even occurs in phenomenological models, where there was an interesting debate between two groups (Pines and Levin) concerning whether such models did or did not generate high $T_c$. Besides the obvious differences of the two phenomenologies (choice of $U_{eff}$, etc.), the main issue concerned how far in energy the dynamic susceptibility extended (Schuttler and Norman, 1996). Most inelastic neutron scattering (INS) studies are confined to less than 100 meV, but we now know that significant weight must be present beyond this energy scale to obtain a high $T_c$. Fortunately, recent INS measurements on underdoped LBCO and YBCO indicate spectral weight up to and beyond 200 meV (for the undoped material, spin fluctuations extend up to 400 meV). On the other hand, the susceptibility decreases with doping, and it is certainly not clear whether there is enough magnetic spectral weight in overdoped materials to be consistent with the relatively high $T_c$ seen.

Of course, arguing about values of $T_c$ might seem analogous to asking how many angels can dance on the head of a pin. After all, in BCS theory, $T_c$ depends exponentially on its coupling constant. Of more relevance is what such theories tell us about experimental data.

Let us start with the one most debated, which is the nature of the phase diagram in cuprates. Three are (of course) a number of versions of this, but in Figure 5 a representative one is given. Besides the well-known magnetic (Néel) insulator at low doping and the superconducting phase at intermediate doping, several other phases have been proposed. Likely, at low temperatures, the Néel state continues as a disordered (spin glass) state, though the range of doping and the ubiquity of this phase is still debated. At high dopings, there is increasing evidence that the normal state is a Fermi liquid, with scattering rates roughly quadratic in temperature and energy, and thus with well defined single-particle (quasiparticle) states. Near optimal doping above $T_c$, one sees a "strange metal" phase characterized by "marginal Fermi liquid" like behavior (Varma *et al.*, 1989). By this, we mean a scattering rate that is linear in temperature and energy. But of most interest is the pseudogap phase in the underdoped regime, first inferred from NMR measurements, and later studied extensively by inelastic neutron scattering, infrared conductivity, tunneling, specific heat, and perhaps most spectacularly by angle resolved photoemission. These studies indicate that an anisotropic gap emerges in the electronic excitations well above $T_c$. The question is what the nature of this gap is.

One thought is that the pseudogap is a precursor to the superconducting gap (Randeria *et al.*, 1992). After all, in almost all magnets, the exchange splitting exists far above the ordering temperature, this temperature being strongly reduced by fluctuations as discussed above. Stated equivalently, the exchange gap is not proportional to $<M>^2$ but rather to $<M^2>$. Superconductors, though, usually do not demonstrate these effects. But cuprates are characterized by small carrier densities, short coherence lengths, and reduced (quasi-2D) dimensionality. All of these conspire to make fluctuation effects more profound. Some spin fluctuation models do advocate that the pseudogap is a pairing gap. But most assume it is actually the magnetic exchange gap itself.

To understand this, note that because of the Mermin-Wagner theorem, long range magnetic order at finite T would not occur in two dimensions. This is why the transition temperature of the undoped material is strongly suppressed relative to the value of J (which is of order 1500 K). That is, $T_c$ is determined by residual three dimensional coupling between the $CuO_2$ planes. But this "three dimensional" $T_c$ is rapidly destroyed by doping. What is left then is a pseudogap state characterized by short range antiferromagnetic fluctuations. As these are known to disappear with overdoping, this provides a natural explanation of the strong doping dependence of the T* (pseudogap) crossover line. In some sense, the pseudogap phase is the "renormalized classical" regime that exists above what is presumably a T=0 magnetic phase transition. The effect is pronounced because of the quasi-two dimensionality (Vilk and Tremblay have demonstrated that this pseudogap is a property of two dimensions and would be very weak in the three dimensional case (Vilk and Tremblay, 1997)).

These approaches have emphasized the potential "quantum critical" nature of the phase diagram shown in Figure 5 (Laughlin *et al.*, 2001). The idea is that at a critical doping, antiferromagnetic fluctuations would disappear (T* would go to zero). This purported "quantum critical point" is buried under the superconducting dome. Is this coincidental? The spin fluctuation proponents say it is not, and note the similarity of the cuprate phase diagram to that determined in a number of heavy fermion magnets. In those cases, the systems are three dimensional, and thus the T* line actually corresponds to the phase line for long range ordering. And in several cases, a superconducting dome appears in the vicinity of where this phase line is going to zero temperature. This was first elucidated by Gil Lonzarich's group for $CeIn_3$ and $CePd_2Si_2$ under pressure (Mathur *et al.*, 1998), but this has now been seen for several other materials as well, including the first known heavy fermion superconductor, $CeCu_2Si_2$. This quantum critical point scenario, though, is not unique to magnetic models.

Despite first appearances, this "nearly antiferromagnetic" picture of the pseudogap phase is quite different from the RVB one to be presented in the next section. In the RVB approach, the fluctuations are singlet in character, but in the spin fluctuation approach, they are antiferromagnetic in nature. Note that a singlet is S=0, but an antiferromagnet corresponds to a mixture of S=0 and S=1, $S_z$=0. They are obviously not the same beast. This controversy is best highlighted by two recent papers, one by Barzykin and Pines (Barzykin and Pines, 2006), the other the RVB review article of Lee, Nagaosa, and Wen (Lee, Nagaosa, Wen, 2006). In both cases, spin susceptibility data are compared to results of the 2D Heisenberg antiferromagnet. In the former case, there is a match (Figure 6), in the latter case there is a large discrepancy. Part of this disagreement is due to the assumed "offset" of the susceptibility: the former assume a temperature independent but doping dependent Fermi liquid component, the latter that the only offset is due to the van Vleck contribution. But the major disagreement concerns the magnitude of J: the latter use the value of 130 meV appropriate for the insulator, but the former assume that T* is actually J itself, and thus strongly doping dependent. To justify this, these authors ironically quote the RVB result that the effective exchange $J_{eff}$ should be J-tx (where t is the hopping and x is the doping).

Of more controversy is the nature of the dynamic spin susceptibility itself. In Figure 7, the famous "hourglass" plot of the energy-momentum relation of the spin fluctuations is presented (Tranqauda *et al.,* 2004). This was first elucidated for underdoped YBCO in its superconducting state (Arai *et al.,* 1999). It is characterized by strong intensity at the neck of the hourglass which occurs at a commensurate wavevector of q=(π,π) known as the resonance, with two incommensurate "bowls" above and below. Although not apparent in Figure 7, it is now known that the hourglass has a 45 degree twist in momentum space, with the incommensurability below resonance oriented along the CuO bond directions and that above resonance oriented along the diagonals (Hayden *et al.,* 2004).

This unusual pattern has been reproduced by various RPA-type calculations. To understand this result, note that in the superconducting state, the polarization bubble becomes $G_k G_{k+q} + F_k F_{k+q}$ where $G_k$ is the normal Greens function and $F_k$ the anomalous (Gor'kov) Greens function. The latter is proportional to the gap, $\Delta_k$. For s-wave superconductors, the presence of a gap causes a $2\Delta$ gap in Im $\chi_0$. But because the gap is constant in the s-wave case, these two terms (GG and FF) destructively interfere in such a way that no pole develops in the RPA expression $\chi = \chi_0/(1 - I\chi_0)$. On the other hand, for the d-wave case, the two terms reinforce near q=(π,π) since the gap product $\Delta_k \Delta_{k+q}$ is negative (Fong *et al.,* 1995). As a consequence Im $\chi_0$ has a step jump at the $2\Delta$ threshold. By Kramers-Kronig, this translates into a log divergence in the real part at ω=2Δ, and thus a pole in $\chi$ is guaranteed at some ω < 2Δ. The dispersion of Im $\chi$ away from (π,π) can either be upwards (magnon-like) or downwards (reverse magnon-like) depending on the Fermi surface geometry. In the latter case, one reproduces the downward part of the hourglass in Figure 7. The upper part of the hourglass is a consequence of the fact that the RPA response is typically incommensurate for frequencies above that of the (π,π) resonance, and some calculations also reproduce the 45 degree twist effect mentioned above (Eremin *et al.,* 2005).

There is, though, an alternate explanation of the data based on "stripes" (Tranquada *et al.,* 2004). The idea is illustrated in Figure 8 (Tranquada *et al.,* 1995). All of the above theories assume homogeneous behavior. But what if instead the system prefers to be inhomogeneous. To understand this, note that each doped hole in a CuO$_2$ plane breaks the four magnetic bonds that connect a given copper ion to its four neighbors. This is energetically costly, and one way to minimize this effect is having the holes clump together. But the long range part of the Coulomb interaction will not prefer this clumping of charge. As a compromise, it was proposed some years ago that the system would instead organize into a lamellar phase where "stripes" composed of the doped holes would be separated by undoped (antiferromagnetic) regions (Zaanen and Gunnarsson, 1989). This idea naturally explains why the separation of the incommensurate wavevector from (π,π) (width of the bottom of the "hourglass" in Figure 7) scales with the doping (the well-known Yamada plot (Fujita *et al.,* 2002) shown in Figure 9). In this picture, the incommensurability is due to the "skip" of the antiferromagnetic structure across the stripe (that is, the antiferromagnetic domains themselves are commensurate), as opposed to RPA, where the incommensurability is due to the 2D Fermi surface geometry (which under certain assumptions (Si *et al.,* 1994) can also reproduce the Yamada plot). And, detailed simulations by several groups have also reproduced the hourglass itself (Uhrig *et al.,* 2004). The upper part of the hourglass is just the gapped magnon-like dispersion one would obtain for an undoped two leg ladder (which is used to model the antiferromagnetic domains). The lower part of the hourglass is due to spin wave-like excitations associated with the stripe periodicity (in the simulations, they are due to the much weaker exchange that couples one spin ladder across a stripe to the next spin ladder). The "twist" of the hourglass also naturally emerges from these simulations.

The important point to emphasize is how different these two explanations of the "hourglass" are. In RPA case, the assumption is a homogeneous 2D material. The spin excitations are derived from underlying fermionic and pair excitations (from $G_k$ and $F_k$). The d-wave symmetry of the gap and the shape of the 2D Fermi surface are crucial for the obtained results. In the stripes case, though, the simulations are essentially undoped spin ladders connected by weak exchange. There are no underlying fermionic degrees of freedom. The physics is crucially dependent on the inhomogeneity of the stripes and their quasi-1D character (the spin gap associated with the upper part of the hourglass is the ladder analogue of the Haldane gap associated with a linear chain of spins). In support of the "stripes" picture, Tranquada's group has seen the hourglass as well for LBCO at x=1/8 (Tranquada *et al.*, 2004) which is not superconducting because of the formation of static stripes (Tranquada *et al.*, 1995). On the other hand, it is quite possible that the RPA-like theories would work in this case as well if the pseudogap had d-wave symmetry. A recent ARPES study has indicated that the T=0 pseudogap state indeed has the same nodal structure as the d-wave superconducting state (Kanigel *et al.*, 2006), and unpublished ARPES results indicate a similar story as well for LBCO at x=1/8 (Valla *et al.*, 2006).

The next experimental controversy concerns unusual features seen in the single particle spectral function measured by ARPES and the resulting density of states measured by tunneling spectroscopy. Both find a very unusual spectral lineshape in the superconducting state (Figure 10), with a sharp peak at the gap energy followed at higher energies by a spectral dip and at even higher energies by a broad hump (peak-dip-hump). It was speculated early on that this might be some strong coupling feature as seen previously in tunneling spectra for conventional superconductors. The idea is that the spectral dip represents a singularity in the electron self-energy. This results in a two branch spectrum, the low energy branch (associated with the sharp peak) represents a renormalized quasiparticle-like dispersion, and a higher energy branch (the "hump") a dispersion which at high energies traces out the bare one (Norman *et al.*, 1997). The theory for this had been worked out in 1963 by Englesberg and Schrieffer in the context of the electron-phonon interaction (Engelsberg and Schrieffer, 1963). The singular energy would then be the sum of the gap energy $\Delta$ and the boson energy $\Omega$ (in their case, a phonon), and thus a subtraction of the gap and dip energies would yield $\Omega$ (the physics of this can be seen from the Feynman diagram in Figure 11). Norman *et al.* noticed that for a slightly overdoped Bi2212 sample the peak-dip-hump structure was only visible below $T_c$; above $T_c$ one simply saw a single broad peak (Figure 10). Based on this, they felt the effect was unlikely to be due to a phonon (which would of course still be present above $T_c$), and more likely due to some kind of electronic collective excitation. Noting that (1) the peak-dip-hump effect was strongest at the $(\pi,0)$ point of the Brillouin zone that are connected to one another by $(\pi,\pi)$ wavevectors, and (2) that the energy of the boson was inferred to be 40 meV, the same as seen for the $(\pi,\pi)$ spin resonance in optimal doped YBCO (the neck of the hourglass in Figure 7), Norman *et al.* speculated that the boson instead was the spin resonance and in a later paper by this group, they gave evidence that the doping dependence of the boson energy was consistent with inelastic neutron scattering, which finds that the resonance energy falls with underdoping (Campuzano *et al.*, 1999). This was very unusual since the gap energy increases strongly as the doping is reduced. Subsequently, Zasadzinski *et al.* (Zasadzinski *et al.*, 2001) traced the mode energy in great detail with tunneling, exploiting its higher energy resolution (Figure 12). They found that the boson energy scales as $5T_c$, just as the resonance does in neutron data. Moreover, with overdoping, the boson energy approaches but never exceeds $2\Delta$, as would be expected for a collective mode inside of a $2\Delta$ gap (as occurs in the RPA calculations discussed above).

With the development of improved (Scienta) detector technology, the momentum and frequency dependence of this effect in ARPES has been mapped out in much greater detail. The peak-dip-hump spectra in constant momentum slices (EDCs) are reflected in a two branch dispersion which is translated to a single "S" shaped dispersion when traced used constant energy slices (MDCs). As with the peak-dip-hump, this "S" shaped anomaly disappears when going above $T_c$ (Sato *et al.*, 2003). All of these effects become less pronounced as one moves in the zone from the $(\pi,0)$ point towards the d-wave node (Kaminski *et al.*, 2001). Along the nodal direction, the MDC dispersion forms a kink behavior (Bogdanov *et al.*, 2000) instead of an "S", and the resulting EDC at the node itself has a "break" (Kaminski *et al.*, 2000) rather than a clear spectral "dip". But the spectral behavior seems to continuously evolve as a function of momentum, indicating that all the strong coupling effects have a similar origin (Kaminski *et al.*, 2001).

This picture, though, has been challenged by a number of groups. Kee *et al* (Kee *et al.*, 2002) have questioned whether there is enough spectral weight in the resonance (typically a few percent of the total spin fluctuation spectral weight) to account for the strong effects seen in ARPES and tunneling. This reduces to an argument concerning the size of the interaction I in the boson spectral function $3/2\ I^2\ \chi(\mathbf{q},\omega)$ (Abanov *et al.*, 2002). More seriously, Lanzara *et al* (Lanzara *et al.*, 2001) have seen the nodal kink in a variety of different cuprates for different dopings at essentially the same energy. They also see a weaker kink-like effect above $T_c$. Because of this, they speculated that the boson was instead a phonon. Extra evidence for phonons was given in later experiments that found an oxygen isotope effect (Gweon *et al.*, 2004) (yet to be reproduced by other groups), and the idea was extended by Cuk *et al* (Cuk *et al.*, 2004) to deal with other regions of the zone (they propose a breathing mode to explain the nodal "kink", and a buckling mode to explain the antinodal "S"). Cuk *et al* advocate that the rapid appearance of the "S" below $T_c$ is due to the gapping of the internal fermion line in the Feynman diagram in Figure 11. Further support for this picture has been given in a recent STM study, where an analysis of the peak-dip-hump structure indicated (1) a doping independent boson energy and (2) a significant oxygen isotope effect (Lee *et al.*, 2006). An advantage of this study is that it was a local probe, and thus the local gap energy (which changes significantly with location in most STM studies of Bi2212) could be subtracted off to determine a local boson energy. A disadvantage of this analysis was that the boson energy was not associated with the dip energy scale as in other studies, but rather with a maximum in the derivative $(d^2I/dV^2)$ spectrum (which corresponds to an energy intermediate between the dip and hump energies). The physical significance of this alternate energy scale is not apparent, since there is no feature in the raw dI/dV data at this energy.

One mysterious fact concerns the doping independence of the nodal kink energy mentioned above. From the Feynman diagram in Figure 11, this energy should be $\Delta+\Omega$ (where $\Delta$ is the gap value at the antinode), but $\Delta$ is known to be a strong function of doping. This requires $\Omega$ to have an opposite doping dependence to compensate. There is no evidence that this occurs for phonons, but this would occur in the spin case for underdoped materials (with $\Omega$ and $\Delta$ have opposite doping dependences (Eschrig, 2006)). A way out is to assume forward (small q) scattering (Kulic and Dolgov, 2005) in Figure 11 instead (so that at the node $\Delta$ reduces to zero), but this is not consistent with the q dependence of the above mentioned phonons (or the spin fluctuations for that matter).

Of course, one could argue that all of these results are suspect in that they implicitly assume Figure 11, which is a lowest order result (and thus inherently assumes a Migdal approximation). It was already mentioned above that such an approximation can be suspect for spin fluctuations. Moreover, some authors have advocated that the hump is a polaron effect, with the hump maximum representing multiple phonon shake offs (Shen *et al.*, 2004). In the spin case, the ultimate strong coupling picture is that advocated by Anderson and co-workers based on the RVB picture (Anderson *et al.*, 2004).

## RVB Theory

This concept was proposed early on by Anderson (Anderson, 1987). Anderson assumed that quantum fluctuations (due to the low spin S=1/2 of the Cu $d^9$ ion and the quasi-two dimensionality of the crystal structure) would melt the Néel order typically associated with Mott insulators. He proposed that the resulting state would be a liquid of spin singlets. As a given singlet involves coupling two of the copper ions, and each copper ion is surrounded by four other copper ions, then each bond can be either part of a singlet or not (Figure 13). Therefore, each bond can "resonate" between being part of a singlet or not, hence the notation "resonating valence bond" (after the work of Pauling, where in benzene, say, one can think of each carbon-carbon link as resonating between a single electron and a double electron bond). Now, imagine doping a hole into such a configuration. Since each copper ion in the undoped case participates in a singlet, then one singlet is broken. This leaves a free chargeless spin (denoted as a "spinon") and a spinless charged hole (denoted as a "holon"). This implies the presence of spin-charge separation. The jury is still out on this particular question. In one dimension, spin-charge separation with its resulting non Fermi liquid characteristics do occur (and is one of the major motivations for those pursuing models based on stripes). But to date, there is no exact answer to this question in two dimensions.

Regardless, Anderson's idea had a profound influence on the field. It emphasized that the cuprates should be thought of as doped Mott insulators, that a "single band" approximation should be adequate, and that non Fermi liquid effects would be pronounced. It also suggested a novel mechanism for superconductivity. Once the doped holes became phase coherent (at temperatures below the phase coherence temperature, which is roughly proportional to the doping), then spin-charge recombination would occur. The resulting charged singlets would be superconducting because of bose condensation of the holons. Although the original prediction for this superconducting state was s-wave, it was realized by several groups within a year that the actual lowest free energy state would be d-wave (Kotliar and Liu, 1988; Zhang *et al.*, 1988).

One interesting prediction of this theory is that the pairing (spin) gap would be maximal at zero doping and then decay approximately linearly with doping (the $J_{eff} = J - tx$ relation mentioned before). On the other hand, the superconducting order parameter would initially be linear in doping, reach a maximum, and then follow the pairing gap for overdoped materials, forming the famous superconducting dome (Figure 14). The resulting phase diagram (also shown in Figure 14) reveals four different states, a superconducting state, a strange metal phase, a Fermi liquid, and a spin gap phase (Nagaosa and Lee, 1992). The RVB spin gap was probably the first prediction for the subsequently observed pseudogap phase. In RVB theory, the pseudogap phase corresponds to a spin singlet state (with its resulting spin gap) but no phase coherence in the charge degrees of freedom. One of the interesting ideas to emerge from this was an explanation for transport in this phase, which reveals a metallic behavior for in-plane conduction, but an insulating behavior for conduction between the planes. In the RVB picture, the metallic behavior is due to the fact that the holons can freely propagate. But to tunnel between the planes, the holons and spinons must recombine to form physical electrons, and this costs the spin gap energy, thus one obtains insulating like behavior for the c-axis conduction (Lee, Nagaosa, Wen, 2006). This "gap" has now been directly seen in c-axis infrared conductivity data (Homes *et al.*, 1993).

At the mean-field level, the RVB physics is relatively well understood. Going beyond mean field theory has been a challenge. One way is to note that when considering a t-J model, the RVB ground state corresponds to a projected BCS wavefunction (the projection designed to remove all $d^{10}$ copper sites, which would be at infinite energy in the infinite U limit). One can perform variational Monte Carlo simulations with such a wavefunction. At zero doping, this state is energetically competitive with the true (Néel) ground state (there is no sign problem in the undoped case, so extensive Monte Carlo simulations have been performed). Though no proof exists for finite doping, we know that the Néel state is rapidly destroyed with doping, and it is anticipated that the RVB state, if anything, will be more competitive once the magnetism disappears. Recent advances in Monte Carlo technology has allowed other quantities besides the group state energy to be calculated. Paremekanti, Randeria, and Trivedi have exploited this to calculate a variety of properties at zero temperature, in particular the doping dependence of the spin gap, the superfluid weight, the quasiparticle residue, the Drude weight, and the nodal Fermi velocity (Paremekanti *et al.*, 2001). The results are in favorable agreement with experiment. Extending these studies to finite temperature and excited state properties remains a challenge for the future. As an aside, one can also perform a partial projection. This is the basis of the gossamer superconductivity theory of Laughlin (Laughlin, 2006), that predicts a ghostly form of superconductivity for small dopings.

The other approach (Lee, Nagaosa, Wen, 2006) has been to exploit the same quantum field theoretic treatments used for heavy fermion materials (the slave boson approach for the Kondo lattice). The spinons and holons are not physical objects, and there is an arbitrary phase relation between the two. This introduces a U(1) gauge degree of freedom in the problem (that cancels for the physical electron). Associated with this is a vector gauge field that acts to satisfy the constraint that the spinon and holon currents balance (the scalar component of the field satisfying the no double occupation constraint). In this formalism, the mean field state is the saddle point of the resulting

Lagrangian. Fluctuations are represented by the quadratic terms associated with the gauge fields. This theory has had some successes (some of the calculated properties qualitatively follow the doping trends found later by the variational Monte Carlo simulations). In particular, it gave the first understanding of how the Fermi surface could be large for hole doped materials (scaling like 1-x) whereas the quasiparticle, Drude, and superfluid spectral weights scale as x (Kotliar, 1995). In essence, the Fermi surface remains large as the doping is reduced, but its spectral weight continuously disappears, so it ends up vanishing, much in the way of the Cheshire cat in Alice in Wonderland. This picture is more or less consistent with photoemission and optics data.

There were problems with the U(1) theory, though. Such a theory predicted that the linear T term in the superconducting penetration depth (due to thermally excited carriers near the d-wave node) scales as $x^2$, which has not been observed. This led Patrick Lee and collaborators to look at an SU(2) generalization (Lee, Nagaosa, Wen, 2006). In the undoped case, the presence of a down spin on a copper site is equivalent to the absence of an up spin (because in the undoped case, every copper site has exactly one spin). As Affleck *et al* point out (Affleck *et al.*, 1988), this implies that the U(1) symmetry previously mentioned expands to an SU(2) symmetry in the undoped case. Connected to this is the fact that in the undoped case, the d-wave spin pairing state is quantum mechanically equivalent to the so-called π flux phase where currents flow around a copper plaquette (Figure 15), these two states being connected by a rotation in particle-hole space. Obviously, this SU(2) symmetry is reduced to U(1) upon the introduction of holes, but the fluctuations implied by this enlarged symmetry group are certainly relevant. In this SU(2) picture, the above mentioned problem is "fixed" (the linear T term in the penetration depth become roughly doping independent (Wen and Lee, 1998)), and in such a formalism, the pseudogap phase can be thought of as a state that fluctuates between a "superconducting" direction and a "flux phase" direction. In other words, the pseudogap phase is a fluctuating mother phase, from which various long range ordered phases emerge at lower temperatures (magnetic and spin glass states at low doping, superconductivity at intermediate doping). Note that the flux phase state is an orbital current phase, and is related to the d-density wave state that has been advocated by others as a phenomenological approach to the pseudogap phase (Chakravarty *et al.*, 2001).

This formalism also gives some idea into the existence of a Nernst effect above $T_c$ (Xu *et al.*, 2000). In normal metals, the Nernst effect (a transverse voltage generated by a thermal gradient) is small due to approximate particle-hole symmetry. But in superconductors, it can be very large in the presence of unpinned vortices (whose flow due to the thermal gradient generates a transverse voltage). Surprisingly, in cuprates, this Nernst effect extends significantly far above $T_c$, implying the existence of vortices well above $T_c$. But we know the superconducting gap is large, so how can such vortices be energetically favorable? In the SU(2) picture, this occurs since the vortex core is the pseudogap phase itself (rather than some gapless normal state) (Lee, Nagaosa, Wen, 2006). This pseudogap, in fact, has been seen in the vortex core of superconducting samples by STM measurements (Renner *et al.*, 1998).

One issue with these types of theories is that they predict certain topological excitations associated with the gauge fields that have yet to be seen by experiment. Until they are, there will always be doubts about such approaches, since they are difficult to employ and the gauge fluctuation expansion is not well controlled, a common bane of strong coupling theories.

On more general grounds, one can ask how different these strong coupling approaches are from the more "weak coupling" approaches discussed earlier. A famous debate has arisen on this subject, with Laughlin claiming that RVB and spin fluctuations represent two different limits of the same underlying theory (Laughlin, 1998), whereas Anderson has strongly differed (Anderson, 1997). As we pointed out earlier, there is one significant difference between these two types of approaches. The spin fluctuation based approaches assume antiferromagnetic fluctuations, whereas in RVB, the fluctuations are singlet in character. It is somewhat surprising that this fundamental distinction has yet to be definitely cleared up by experiment.

## Alternate Mechanisms

Space prohibits a detailed summary of the countless theories that have been proposed in the context of cuprate superconductors. But in this section, those of some note will be mentioned, especially in connection to what was discussed above. This section (and article) is then ended with a brief discussion of the "phonon" question.

The SO(5) approach of Zhang and collaborators (Demler $et$ $al.$, 2004) is similar in spirit to the SU(2) approach just mentioned. Instead of fluctuating between a flux phase and a superconducting phase, one fluctuates between the two known ground states (antiferromagnetism and superconductivity). The minimal group which contains these two order parameters is SO(5) (the "five" being the real and imaginary values of the superconducting order parameter, and the three spatial components of the Néel vector). SO(5) has ten generators, four of them being the charge operator and the three spin components, the other six are so-called $\pi$ operators that connect the superconducting and Néel sectors of the theory. These operators, acting on the superconducting ground state, create the previously mentioned $(\pi,\pi)$ resonance - that is, this resonance can be thought of as an excited triplet pair state with center of mass momentum $(\pi,\pi)$. In the SO(5) case, though, this resonance is a property of the particle-particle channel, and only appears in neutron scattering because of particle-hole mixing in the superconducting state. This theory thus naturally explains why the resonance is only seen below $T_c$, and why its doping and temperature dependences scale with the superconducting order parameter (a property which does not obviously follow from the RPA calculations previously discussed). It also predicts that the vortex core is antiferromagnetic (Arovas $et$ $al.$, 1997) and that charge modulation effects seen in tunneling for underdoped samples are from a checkerboard pair density wave state (Chen $et$ $al.$, 2002) (as opposed to stripes). There are some issues connected with this theory, though. The $(\pi,\pi)$ resonance in this theory is an antibound state, as opposed to RPA where it is a bound state. Available data are much more consistent with the latter, as the resonance energy is less than $2\Delta$ (that is, it lies below the particle-hole continuum rather than above (Tchernyshyov $et$ $al.$, 2001)). Moreover, although the theory incorporates the fact that the undoped material is an antiferromagnet, it does not take into account the fact that it is a Mott insulator (where the charge excitations are strongly gapped). To correct this obvious deficiency, a modified theory known as projected SO(5) has been developed, and the reader is referred to the literature for a discussion of this technique and how it addresses spectroscopy data like photoemission in the low doping regime (Zacher *et al.*, 2000). Space also prohibits a discussion of other "preformed" pairs scenarios, such as the $QED_3$ theory of Franz and Tesanovic that advocates that the pseudogap phase is a phase disordered superconductor characterized by a proliferation of vortex-like excitations (Franz and Tesanovic, 2001).

The $\pi$ flux phase mentioned in the context of RVB theories is an orbital current phase with an associated wave vector of $(\pi,\pi)$. Such a phase is characterized by point nodes, that have been recently inferred as the T=0 ground state of the pseudogap phase by thermal conductivity (Sutherland *et al.*, 2005) and more recently photoemission data (the latter directly imaging the nodes (Kanigel *et al.*, 2006)). On the other hand, there is no evidence from photoemission that the pseudogap (at least the low energy one associated with the leading edge of the ARPES spectrum) has a finite q vector associated with it. In fact, current ARPES data are consistent with the pseudogap being tied to both the Fermi surface and to the Fermi energy, as would be expected for a q=0 state (a superconductor is a q=0 state, since the center of mass momentum of the pair is zero). But the $\pi$ flux phase state is not the only orbital current phase that has been proposed. Varma has argued that when one reduces from the three band model (copper $dx^2$-$y^2$ orbital and oxygen $p_x$ and $p_y$ orbitals) to the commonly employed single band model (the antibonding mixture of the copper and oxygen states), one has thrown out the baby with the bathwater so to speak (Varma, 2006). He believes a complete theory must keep all of these degrees of freedom. Although a discussion of this important topic would take us far outside the bounds of this review, it should be noted that optics data have revealed changes in spectrum when passing through $T_c$ which extend up to several eV (Rubhausen *et al.*, 2001). If these degrees of freedom are important for superconductivity, then indeed neglecting higher energy degrees of freedom could be dangerous. In Varma's theory, a unique orbital current phase emerges because of a non-trivial (Berry) phase involving the three different orbitals. This shows up in the antibonding band at the Fermi energy as an orbital current which flows in the sub plaquette formed by a copper ion and its surrounding oxygen neighbors (Figure 16). Since the structure is based on the unit cell itself, it is a q=0 state. Originally, the theory had a current pattern such that would generate magnetic reflections for diagonal aligned Bragg vectors (right panel, Figure 16). These were searched for by neutron scattering and not found. A related circular dichroism experiment was performed in photoemission by Kaminski *et al* (Kaminski *et al.*, 2002) and found a dichroism shift along the $(\pi,0)$-$(\pi,\pi)$ direction (not yet reproduced by other groups, though). To account for this, Varma rotated his current pattern by 45 degrees (left panel, Figure 16). The resulting magnetic reflections would then be along the bond directions, and these were recently seen by neutron scattering (Fauque *et al.*, 2006) (but again, not reproduced by other groups). Varma's theory also predicts nodes in the pseudogap phase, as has been recently inferred from experiment (as mentioned above). Recently, a dichroism signal has been found by x-rays that matches the ARPES one (Kubota *et al.*, 2006), but the claim was that the signal was structural, not magnetic, in origin.

Another inhomogeneous pattern, as mentioned before, is stripes. These patterns have been seen in several transition metal oxides, and in the LTT structural phase of doped lanthanum cuprate near x=1/8 (Tranquada *et al.*, 1995). As mentioned before, stripes give a natural explanation of the Yamada plot, have been proposed (in their dynamic version) as an explanation for the unusual hourglass shaped spin dispersion seen in neutron scattering, and can account for various Fourier charge peaks seen in STM data (Kivelson *et al.*, 2003). From a theoretical perspective, the advantage is that the quasi-1D nature of the stripes naturally yields spin-charge separation and its resulting non Fermi liquid properties (which are more difficult to generate in homogeneous 2D models). The holes pick up their pairing gap by virtually hopping from the stripes to the undoped antiferromagnetic domains (Emery *et al.*, 1997) (even leg spin ladders having a spin gap). The much lower value of $T_c$ as compared to the gap is determined by Josephson coupling of the stripes. The stripes picture has been important in focusing the physics community on the fundamental question of real space based approaches as compared to the traditional momentum space based approaches used in the past to address superconductivity. It emphasizes the role of inhomogeneity, which has been spectacularly seen in STM experiments (Pan *et al.*, 2001). It provides a unique (quasi-1D) approach to the cuprate problem. But this approach has also raised a number of questions. A recent x-ray analysis of the x=1/8 stripe phase in LBCO has found a smooth sine wave for the charge modulations, as opposed to the square wave picture of Figure 8 (Abbamonte *et al.*, 2005). Charge modulation effects seen in STM are weak in intensity, and moreover tend to trace out a checkerboard pattern (Hanaguri *et al.*, 2004) as opposed to a stripe one (a checkerboard has been advocated for the spin pattern in doped LSCO as well in a recent neutron scattering study (Christensen *et al.*, 2004)). Even the large inhomogeneity effects seen by STM have been challenged by others (the group of Oystein Fischer typically does not observe them). Alternate explanations have also been given concerning the STM Fourier peaks (McElroy *et al.*, 2003) (for instance, similar ARPES autocorrelation studies are consistent with a joint density of states explanation for the Fourier peaks (Chatterjee *et al.*, 2006)). But an interesting aspect of the stripes scenario is the possible coupling of lattice and charge/spin degrees of freedom. It is well known that certain phonons show anomalies that are thought to be connected with static stripes or their dynamic variants (Reznik *et al.*, 2006).

Having mentioned the lattice, it is time to end this review by a discussion of the "phonon" question. As mentioned earlier, it has been advocated that the peak-dip-hump lineshape seen in tunneling and ARPES is a strong polaron effect. Polarons are certainly prominent in other transition metal oxides, such as manganites, and in fact were an integral part of the guiding principle that led Bednorz and Mueller to their original discovery. Whether they are present at optimal doping is an entirely different matter (the normal state actually being a quite good metal with well developed screening as characterized by the 1 eV plasmon). And, as mentioned before, it has been advocated by several photoemission and STM groups that the strong coupling anomalies seen in those spectra are caused by phonons rather than magnetic excitations.

Of course, the one known thing in superconductivity is that phonons can definitely cause pairing. All finite frequency phonons contribute positively to the s-wave pairing channel (Bergmann and Rainer, 1973). Obviously, only some of them do for the d-wave channel. An advantage of the d-wave channel is the strong reduction in the direct Coulomb repulsion (due to the nodes in the pair wavefunction), but then again, we don't know of any s-wave electron-phonon superconductors which occur at 150K, much less in the d-wave channel with its reduced coupling constant. So, it is a rather far stretch to believe that phonons can account for cuprate superconductivity. This does not mean that they cannot be responsible for certain anomalies in experimental spectra, and even those authors advocating such are careful not to claim that phonons are solely responsible for superconductivity. But what is somewhat disturbing is the trend to fit the entire spectrum assuming phonons (or polarons) and ignore the underlying strong electron-electron interactions that presumably give rise to the various states (Mott insulator, d-wave superconductor) to begin with. It is difficult, of course, to properly treat all degrees of freedom (charge, spin, lattice), particularly for doped systems, and then one is faced with the "everything but the kitchen sink" scenario for describing the material. In $^3$He, it is known in fact that many degrees of freedom enter the various interactions, including the pairing one. But there, a spin fluctuation based approach captures the fundamental essence of the problem. It is quite likely that such a spin fluctuation approach (or a strong coupling analogue like RVB) will also capture the essence of the cuprate problem. But only time will tell.

## Acknowledgmenents

The author would like to thank many people for discussions that led to his thoughts as laid down in this review. A far from exhaustive list would include Phil Anderson, J C Campuzano, Andrey Chubukov, Seamus Davis, Matthias Eschrig, Steve Kivelson, Alessandra Lanzara, Bob Laughlin, David Pines, Mohit Randeria, Doug Scalapino, Z-X Shen, and Chandra Varma. The author acknowledges support from the US Department of Energy, Office of Science, under Contract No. W-31-109-ENG-38.

**Figures**

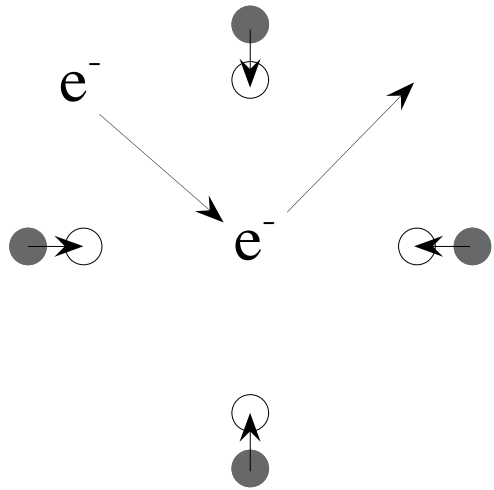

Figure 1 – The electron-phonon interaction leads to an induced attraction between electrons. Arrows joining circles show displaced ions; the time scale of these ions for relaxation back to their original positions is slow compared to the electron dynamics.

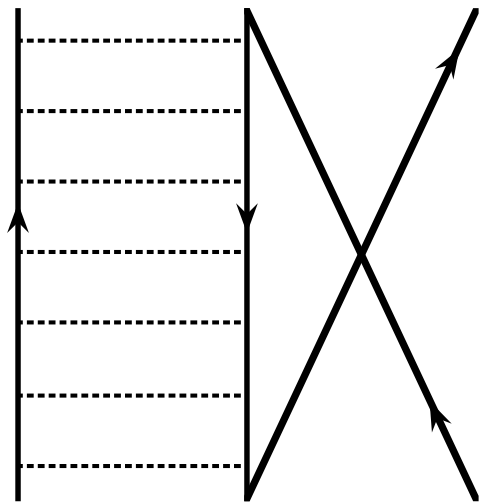

Figure 2 - Particle-particle interaction from spin fluctuations. Note the particle-hole ladder sum buried inside this diagram.

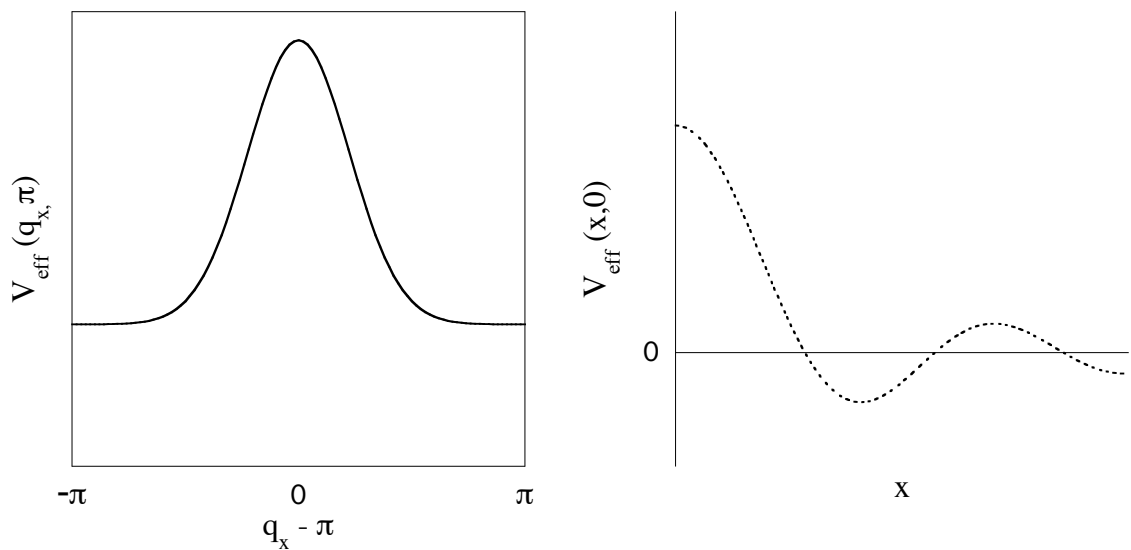

Figure 3 - Effective interaction for spin fluctuation mediated pairing (antiferromagnetic case). Left panel is for momentum space (a repulsive potential peaked at q=(π,π)), right panel for real space (repulsive potential on-site, attractive potential for a near neighbor separation).

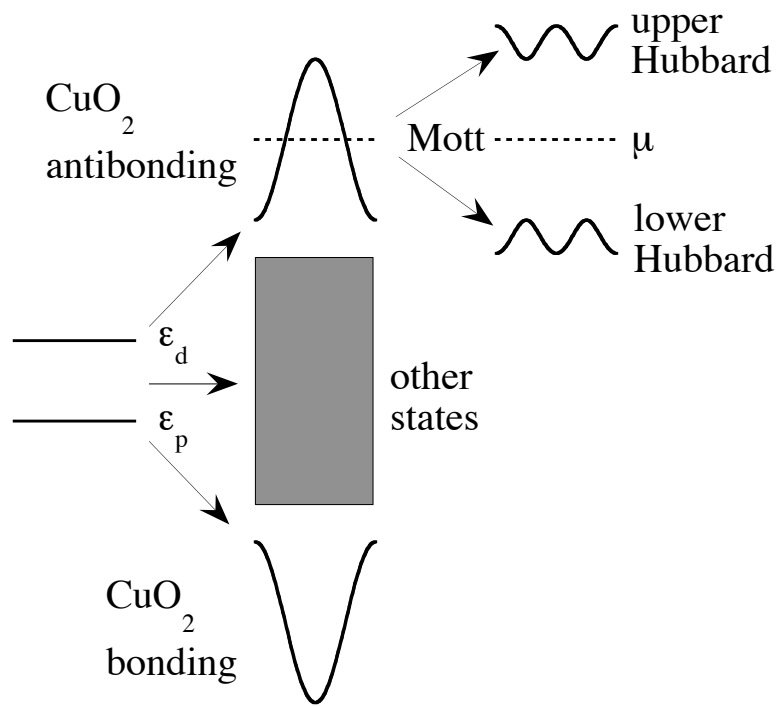

Figure 4 - Electronic structure of the layered cuprates. The copper d and oxygen p levels hybridize, resulting in a partially filled antibonding band. A Mott gap due to Coulomb correlations splits this band, leading to the formation of an upper Hubbard band and a lower Hubbard band, with the chemical potential, μ, inside this gap for zero doping.

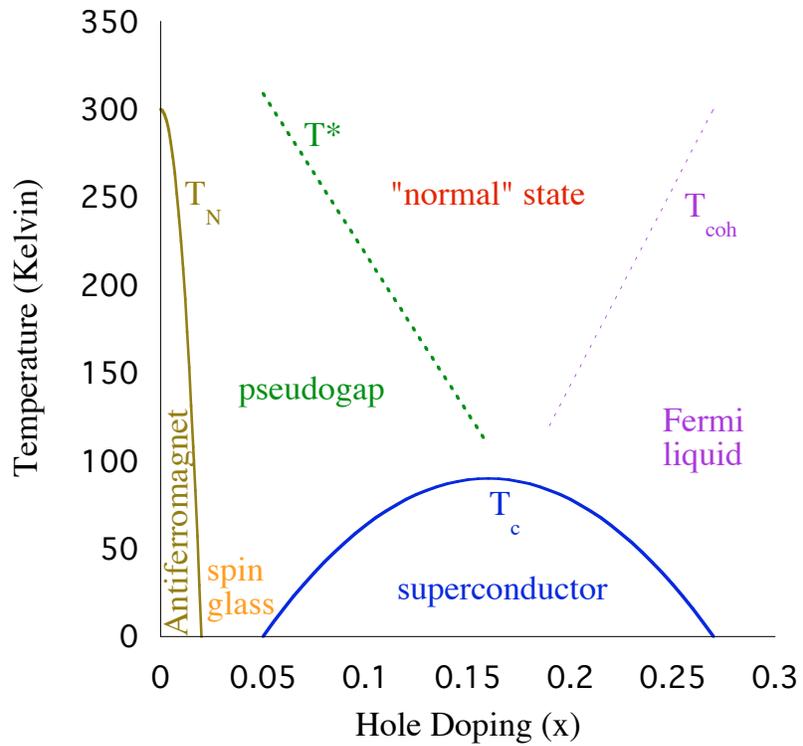

Figure 5 - Phase diagram of the cuprates versus x, the number of doped holes per copper ion. Solid lines represent true thermodynamic phase transitions; dotted lines indicate crossover behavior.

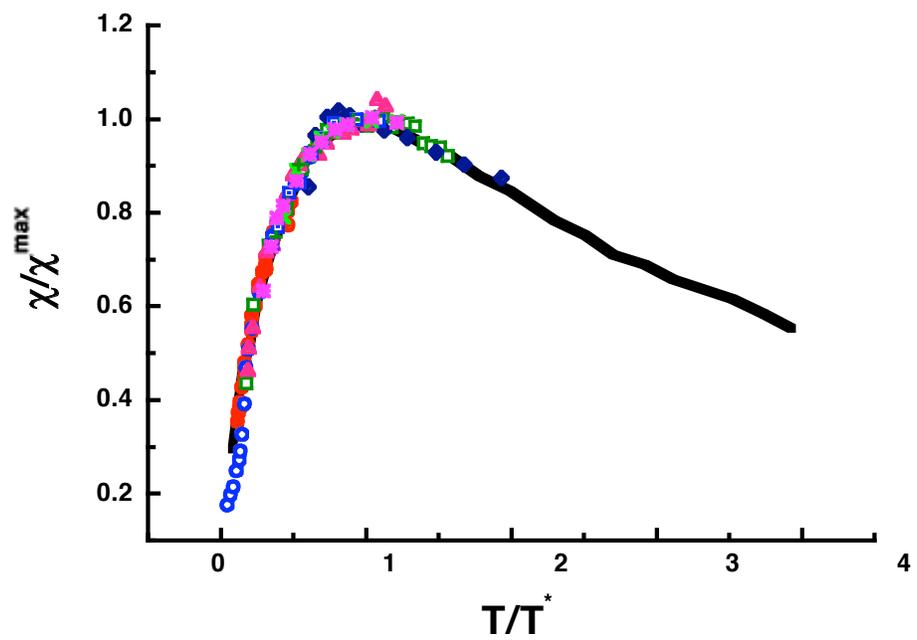

Figure 6 – Comparison of the 2D Heisenberg model to experimental susceptibility data in YBCO (Barzykin and Pines, 2006). The data are scaled assuming that the effective exchange at a given doping is equal to the pseudogap onset temperature T*.

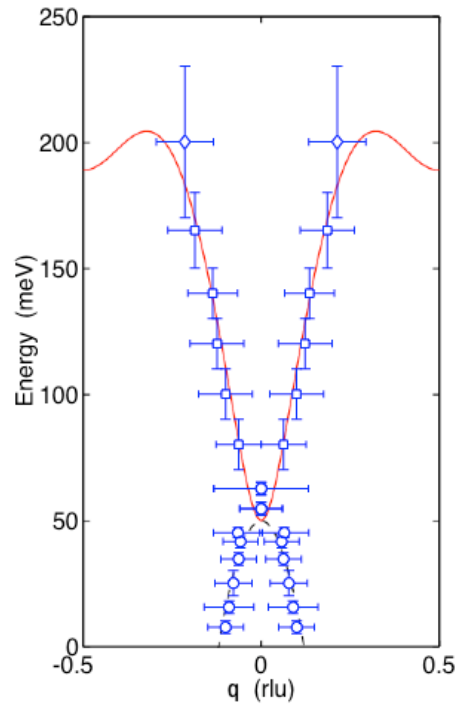

Figure 7 - Hourglass like dispersion of the spin excitations in the stripe ordered phase of LBCO, measured with respect to q=(π,π), as revealed by inelastic neutron scattering (Tranquada *et al.*, 2004). A similar dispersion is seen in the superconducting state of underdoped YBCO (Hayden *et al.*, 2004).

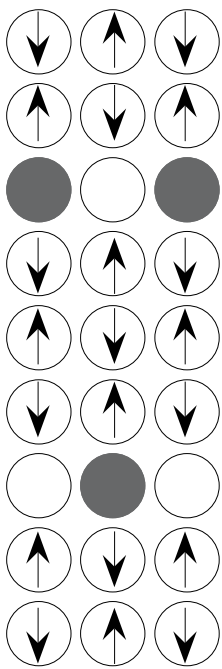

Figure 8 - Stripe picture for x=1/8 doping - circles are copper sites, arrows represent spins, dark circles doped holes.

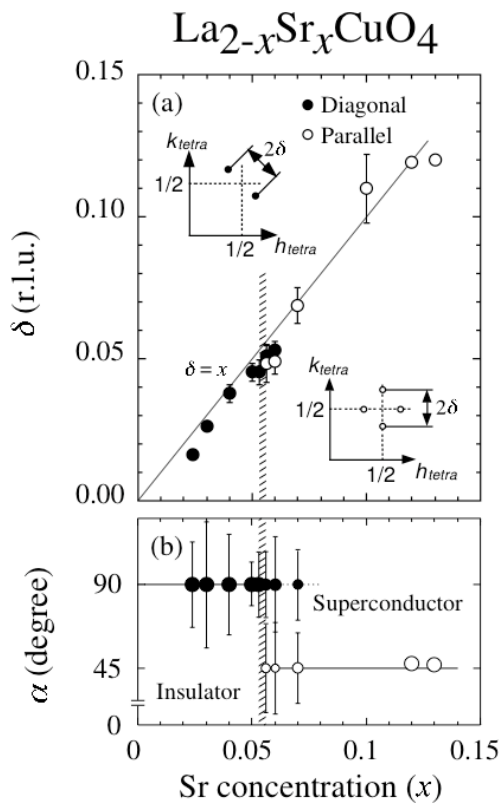

Figure 9 - Neutron scattering peaks versus doping for LSCO. The spot pattern rotates by 45 degrees at the spin glass/superconducting boundary. δ is the incommensurability, and α the angle of the spots in momentum space relative to the (π,π) wavevector (Fujita *et al.*, 2002).

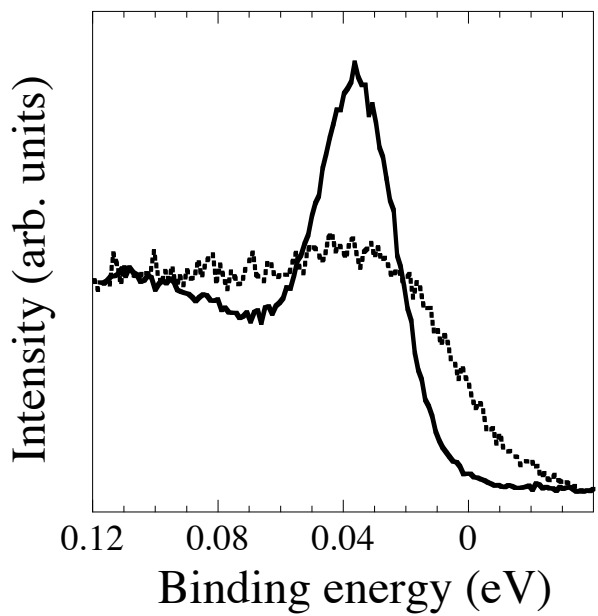

Figure 10 - ARPES spectra at k=(π,0) for an overdoped ($T_c$=87K) Bi2212 sample in the normal state (dotted line) and superconducting state (solid line).

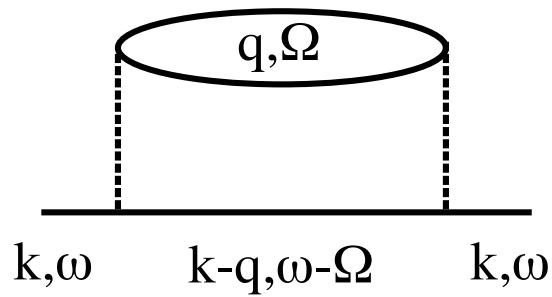

Figure 11 - Lowest order Feynman diagram for electron-electron scattering. For the spin resonance case, the bubble labeled by (q,Ω) is replaced by the dynamic spin susceptibility. For the phonon case, it is replaced by the phonon propagator.

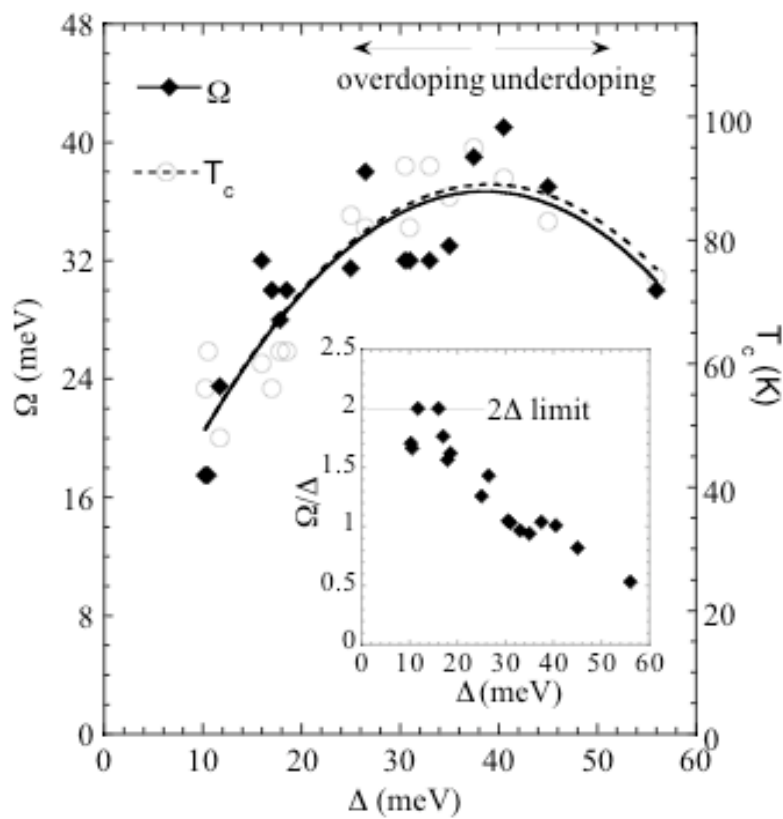

Figure 12 - Resonance energy versus doping inferred from tunneling data, showing scaling with $T_c$. The inset demonstrates that this energy saturates to $2\Delta$ in the overdoped limit (Zasadzinski *et al.*, 2001).

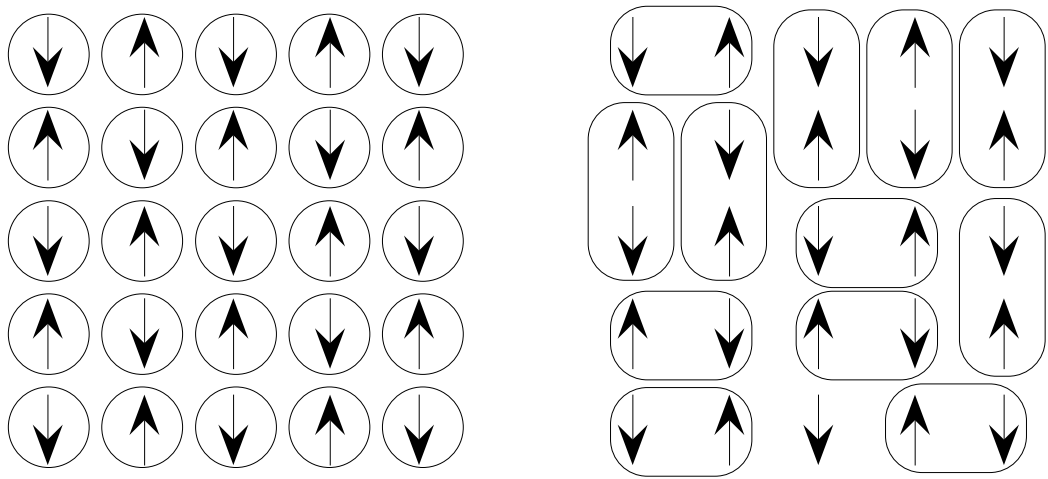

Figure 13 - Néel lattice (left panel) versus RVB (right panel). The RVB state is a liquid of spin singlets. Circles are copper sites, arrows represent spins, ovals spin singlets.

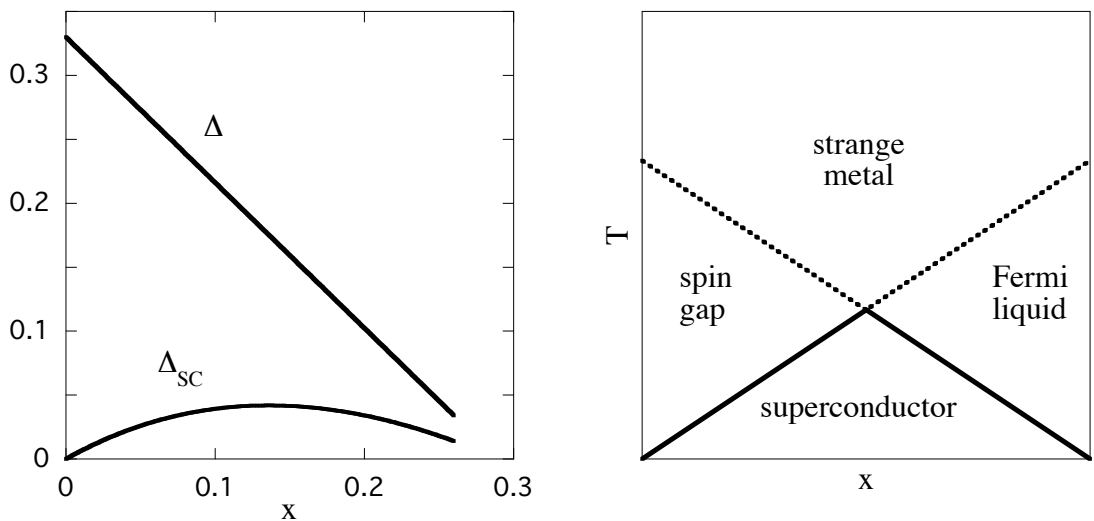

Figure 14 - Spin gap, $\Delta$, and superconducting order parameter, $\Delta_{SC}$, as a function of doping from RVB theory (left panel). RVB phase diagram (right panel).

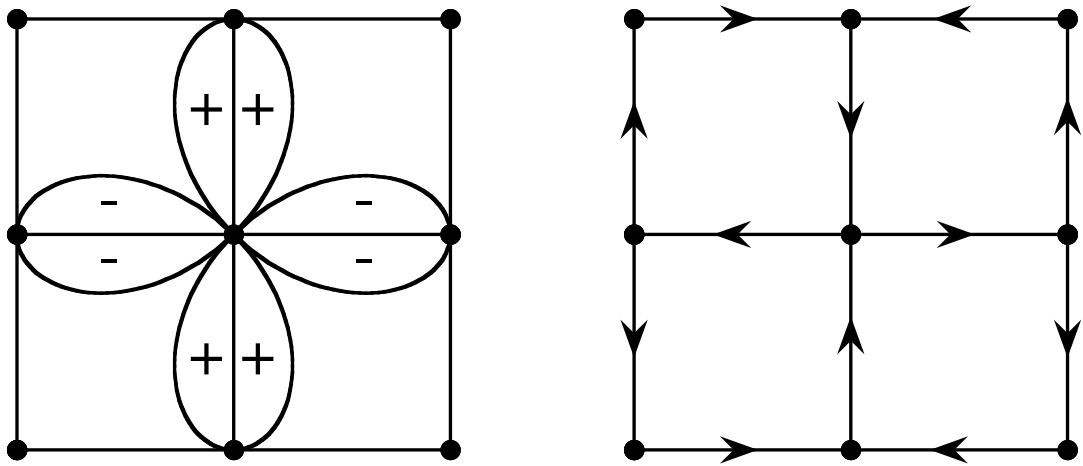

Figure 15 – Two RVB states that are equivalent at zero doping: d-wave spin pairing state (left panel) and π flux state (right panel). Dots are copper ions, arrows are bond currents.

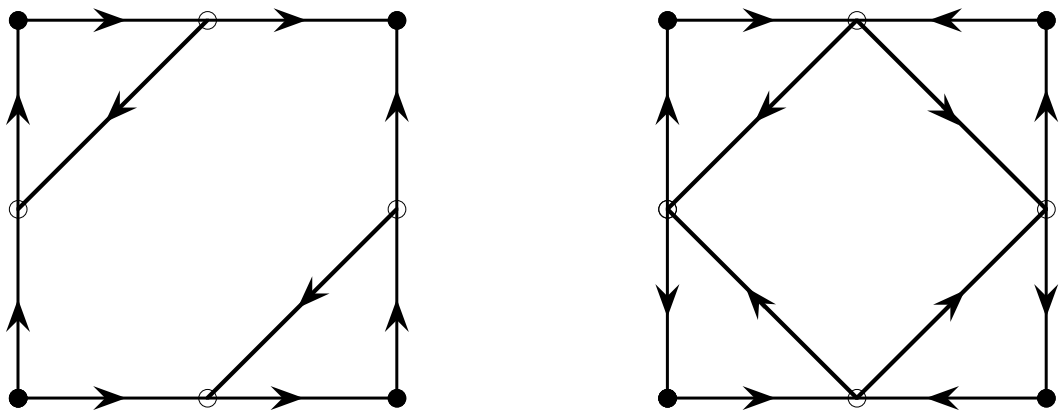

Figure 16 – Two orbital current patterns proposed by Varma. Solid dots are copper ions, open dots oxygen ions, and arrows are bond currents. The left pattern has been used to interpret recent polarized ARPES and neutron results in the pseudogap phase.